\documentclass[10pt,emptycopyrightspace]{ewsn-proc}

\usepackage{graphicx}
\usepackage{balance}
\usepackage{comment}

\usepackage{cite}
\usepackage{hyperref}
\usepackage{orcidlink}
\usepackage{comment}
\usepackage{amsmath,amssymb,amsfonts}
\usepackage{algorithmic}
\usepackage{graphicx}
\usepackage{textcomp}
\usepackage{academicons}
\usepackage{xcolor}
\usepackage{algorithm}
\usepackage{algorithmic}

\numberofauthors{1}
\author{\\
\alignauthor Jimmy Fernandez Landivar$^1$\orcidlink{0000-0002-4904-5256}, Pieter Crombez$^2$, Sofie Pollin$^1$\orcidlink{0000-0002-1470-2076} and Hazem Sallouha$^1$\orcidlink{0000-0002-1288-1023}\\
   \affaddr{$^1$Departement of Electrical Engineering (ESAT) - WaveCoRE}, KU Leuven, Belgium\\
   \affaddr{$^2$Televic Healthcare, Belgium}\\
   \email{\normalsize{jfernand@esat.kuleuven.be, p.crombez@televic.com, sofie.pollin@kuleuven.be, hazem.sallouha@kuleuven.be }}
}

\title{QualityBLE: A QoS Aware Implementation for BLE Mesh Networks
}

\begin{document}

\maketitle

\begin{abstract}

Bluetooth Low Energy (BLE) Mesh is widely recognized as a driver technology for IoT applications. However, the lack of quality of service (QoS) in BLE Mesh, represented by packet prioritization, significantly limits its potential. This work implements a quality-of-service (QoS) method for BLE Mesh to prioritize the data packets and provide them with different network transmission settings according to their assigned priority. Unlike existing works on QoS for BLE Mesh, our proposed approach does not require any modifications to the BLE Mesh protocol and can be smoothly adopted in existing BLE Mesh networks. We conducted an extensive measurement campaign to evaluate our solution over a 15-node BLE Mesh network deployed to emulate a smart healthcare scenario where 45 sensors with an assigned priority transmit information over the network. The experiments provide performance results for single and multi channel network scenarios. The obtained results validate our solution, showing the difference between the established priorities and providing insights and guidelines to conduct further research on QoS over BLE Mesh and broadcast-based networks.

\end{abstract}

%
%


\section{Introduction}
  \label{Sec_I}
The Internet of Things (IoT) applications space is constantly increasing, covering many areas, from home, transportation, and agriculture to healthcare and industry \cite{meshintro,sallouha2017ulora}. As a result, the number of IoT devices is rising, setting a diverse range of requirements on quality-of-service (QoS), power consumption, coverage, and spectral efficiency \cite{sallouha2017ulora}. Bluetooth Low Energy (BLE) is one of the most popular IoT technologies which suffices IoT applications with a limited number of nodes. In order to scale up BLE networks, a multi-hop-based mesh BLE standard was released in 2017\cite{Darroudi2017}. 

The BLE Mesh standard, which is built on top of the BLE lower protocol layers, uses BLE's advertising/scanning states to broadcast messages to other devices, allowing them to communicate in a mesh structure \cite{Darroudi2017}. A BLE Mesh network can theoretically include up to 32767 devices, and a BLE Mesh packet can perform up to 127 hops to reach its destination \cite{mesh101}. These characteristics positioned BLE Mesh as a favorable technology for several applications where network power consumption and scalability, in terms of the number of devices and range, are more critical than latency \cite{Sergi2021meshiot}. These applications include home automation \cite{Darroudi2017}, healthcare \cite{Sergi2021meshiot}, and smart infrastructures \cite{Aijaz2021why}. While BLE Mesh meets the connectivity requirements of these applications, it sends all packets without any prioritization, failing to provide any QoS needed in modern IoT networks. For instance, in modern smart hospitals where patients have wearable devices, transmitting important vital signs or nurse calls must have a higher priority than an asset tag that sends a control signal or a command to turn on the illumination system.

The use of IoT in the area of healthcare applications has recently attracted considerable research focus. The work presented in \cite{Nausheen2018} showed the benefits of IoT in healthcare, highlighting the necessity of supporting multiple types of packet priorities in the network in big health infrastructures such as hospitals. In \cite{Sergi2021meshiot}, the authors demonstrated IoT networks' essential role in monitoring applications in general and for healthcare in particular. However, the work was focused on BLE implementations only and did not address the scalability problem faced in massive deployments such as in the case of hospitals. The urgent need for a QoS implementation that can be integrated in the BLE Mesh standard motivates our work in this paper, which is, to the best of our knowledge, an open research question. 

This paper proposes a QoS implementation for BLE Mesh networks, which can be integrated into the network while staying in line with the BLE Mesh standard. In particular, our QoS implementation uses only 1-Byte of the packet sequence number (SEQ) field on the network Protocol Data Unit (PDU) structure to incorporate a priority class to every packet. Different network transmission parameters are automatically set for each priority on the nodes.
We evaluate the performance of our solution in a 15-node BLE Mesh testbed deployed in a full building floor, emulating a smart healthcare infrastructure use case. Our main contributions are summarized as follows:
\begin{itemize}
\item We introduce a QoS implementation method for the BLE Mesh protocol stack. Specifically, this method adds a priority class (up to 256 values) to the network layer PDU. It is associated with various network transmission configurations that modify the Transmission Power, Time to Live (TTL), Number of rebroadcasts, and packet Advertising Interval. This implementation enables the possibility to differentiate the network transmission services for every priority class.
\item An extensive measurements campaign is performed on a 15-node (nRF52 System-On-Chip and Zephyr RTOS-based) BLE Mesh network. Each of our nodes virtualizes three sensors, representing a 45-sensor network in total.

\end{itemize}

The rest of this paper is organized as follows: Section \ref{Sec_I} defines the related work and state-of-the-art. Section \ref{Sec_II} presents the BLE Mesh networking process.
Section \ref{Sec_III} describes the proposed method for QoS implementation in BLE Mesh. Section \ref{Sec_IV} details the data collection and experiment campaign. Finally, sections \ref{Sec_V} and \ref{Sec_VI} present the performance evaluation results, conclusions, and future work.

\section{Related Work}
\label{Sec_I}

Since the BLE Mesh standard's official release, several works have studied its performance regarding scalability, reliability, and latency. The author in \cite{Murillo2019} implemented a low-cost BLE Mesh testbed with multiple configurable nodes, including the possibility to test flooding and route-based mesh protocols. The work in \cite{sillabs} presented the results of massive experimentation on BLE Mesh, showing results related to the network performance, latency, and size in an environment with coexistence of other wireless technologies. The performance of a BLE Mesh network composed of heterogeneous devices deployed in an area of 1100 m\textsuperscript{2} was tested and analyzed in \cite{king}. The work in \cite{Giacomini2020} suggested methods to improve BLE Mesh and its node features by applying self-organizing network concepts to the BLE Mesh node roles. The authors in \cite{challenges} suggested deployment strategies and node configuration parameters in order to improve the performance of BLE Mesh. However, none of these aforementioned works introduced a QoS solution for BLE Mesh network as they mainly focused on a signle traffic class.
Only a few existing works addressed the QoS problem in BLE Mesh. These works either require frequent manual configurations \cite{Rondon2020}, propose different strategies for different network configurations \cite{whitepaper}, or make changes on the inner default functions on the protocol itself \cite{Basu2021}. Moreover, a framework for introducing QoS in BLE Mesh by varying some features of the radio standard, such as the backoff time mechanism and channel selection, is proposed in \cite{Basu2021}. However, solutions modifying critical elements in the standard limit the possibility of their adoption on a large scale, where following a standard protocol is a key requirement. The urgent need for a QoS implementation in BLE Mesh networks, that stays in line with the technology protocol motivates our work in this paper.

\section{BLE Mesh Background}\label{Sec_II}

This section presents the principal characteristics of the BLE Mesh communication process, network architecture, and protocol stack.

\subsection{Elements, models, and node roles}

Packets in BLE Mesh are sent using a controlled flooding protocol where the nodes broadcast the packets using BLE's advertising and scanning channels. BLE Mesh nodes receive the packets and rebroadcast them until reaching the destination or traverse all nodes in the mesh network. This mechanism is called \textit{relaying} and allows the packets to do multiple hops between devices by being \textit{relayed}. The number of hops a packet can do is controlled by a parameter called Time to Live (TTL) \cite{meshintro,Darroudi2017}. The TTL decreases with every hop made by the packet, limiting the number of hops a BLE Mesh packet can perform. When TTL reaches 0, the packet will not be relayed \cite{meshintro}.

\begin{figure}[t]
\centering
\includegraphics[scale=0.175]{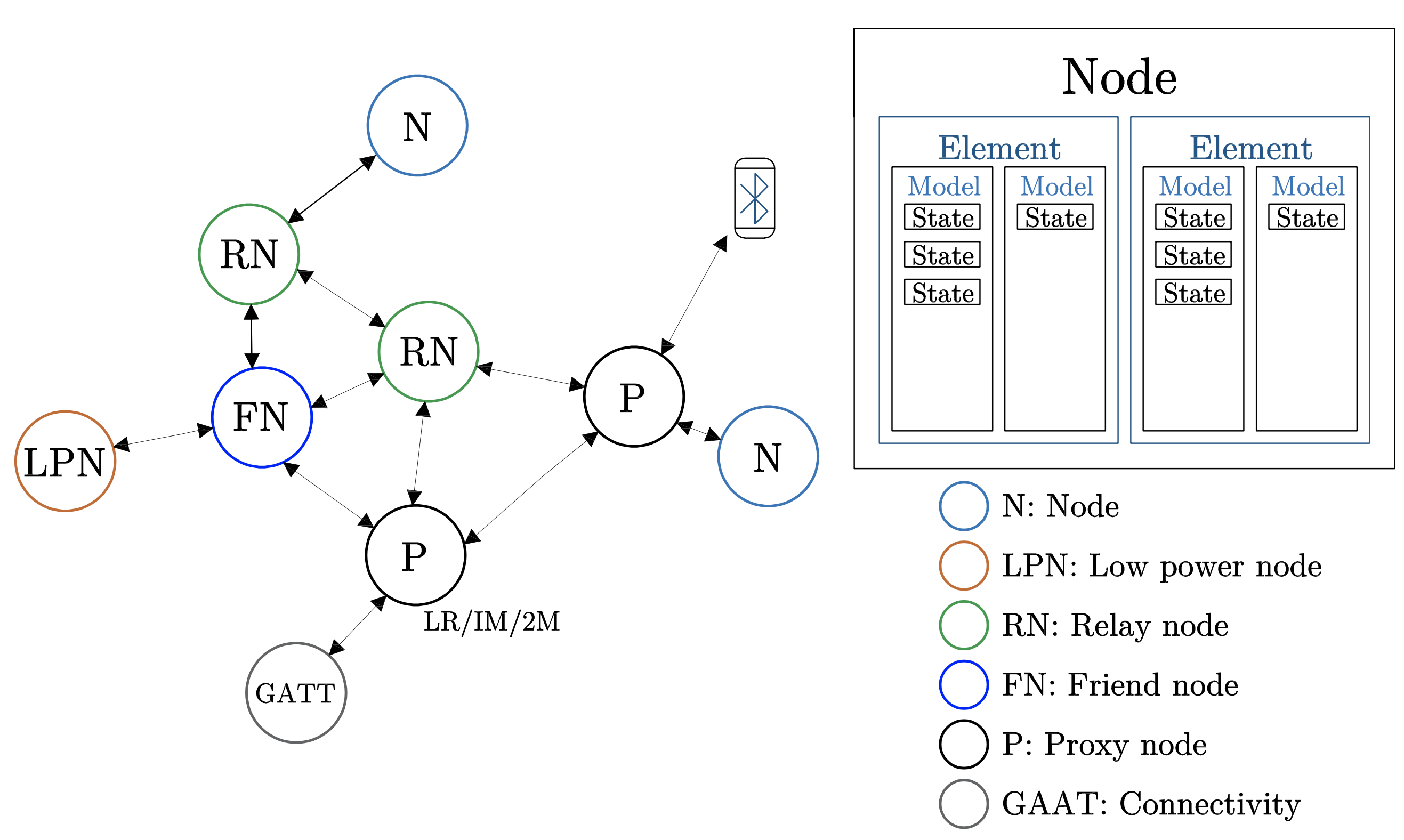}
\caption{BLE Mesh network architecture.}
\label{fig:meshnodes}
\vspace{-1em}
\end{figure}

The BLE Mesh network integrates different types of nodes that can send, relay, and receive messages \cite{meshintro}. The standard defines four types based on their role: First, \textit{The Relay Nodes} can receive and re-transmit messages. Second, \textit{The Proxy Nodes} enables the communication between BLE devices and the mesh network. Third, \textit{ The Low Power (LP) Nodes} are in energy-saving mode until they awake and communicate with the mesh network through the Friend nodes. Fourth, \textit{The Friend Nodes} enables the communication between the LP nodes and the mesh network, receiving and sending information collected from the LP nodes and relaying it to the other nodes. Every BLE Mesh node supports the four roles, and they can be set by software according to the network design. Figure \ref{fig:meshnodes} depicts the BLE Mesh network architecture.

The BLE Mesh nodes can have one or more roles, and every node has a unicast address. In their architecture, the nodes can have multiple constituent parts that establish communication with the mesh network jointly or independently. These parts are called \textit{elements}\cite{Woolley2019model}. A node can have one or more elements, and every element has a different unicast address. A group of nodes including different elements can have a group address\cite{meshintro}. The functions of every element to interact with the mesh network are defined by a structure called a \textit{model}\cite{Woolley2019model,mesh101}, as illustrated in Figure \ref{fig:meshnodes}. The models communicate with different nodes using mesh messages. The messages are part of the BLE Mesh packet and are transmitted by nodes from and to an address. These addresses could be unicast or group addresses\cite{meshintro}. A node can only send messages to an address it has been registered to publish and receive them only if it has subscribed. Each message is defined by a unique \textit{opcode} that allows communication between nodes in the mesh network, identifying the type of information and assigning a model to process it\cite{Woolley2019model}. 

\subsection{BLE Mesh protocol stack architecture}

The BLE Mesh protocol comprises seven layers on top of the BLE protocol stack. Here, the BLE's physical and link layers establish radio communication through advertising and scanning channels 37, 38, and 39 \cite{mesh101}. From the bottom to the top, the seven different layers and their functions are:
\begin{enumerate}
    \item \textit{The bearer layer} composed of the advertising bearer and the GATT bearer. The advertising bearer is the one that uses the BLE functions of scanning and advertising to send and receive packets exclusively for BLE Mesh.
    \item \textit{The Network Layer} is in charge of sending the encrypted transport PDUs to the bearer layer, adding the addresses and extra packet information to be sent. 
    The relay process is carried out here. If a packet is received and if it belongs to the node, it goes to the upper layer; otherwise, it is discarded or relayed.
    \item \textit{The Lower Transport Layer} sends the PDU to the lower transport layer in the peer devices. Here, the data is segmented if necessary.
    \item \textit{The Upper Transport Layer} is where the application data is encapsulated and encrypted to be sent to the lower transport layer. 
    \item \textit{The Access Layer} verifies the data to be encrypted and sent it to the transport layer. It also verifies that the data coming from the transport layer is correct to be sent to the upper foundation models' and models' layers. 
    \item \textit{The Foundation Models' Layer} is where the models and their different states are interpreted and defined. 
    \item \textit{The Models Layer} is the one that associates the different models and states with the user applications.
\end{enumerate}

\section{QoS Implementation for BLE Mesh}\label{Sec_III}

In this section, we introduce our novel QoS implementation for BLE Mesh to enable priority and differentiate services for the data packets flooding the network. Our method is performed over the nRFConnectSDK\footnote{https://www.nordicsemi.com/} BLE Mesh protocol stack of the Zephyr RTOS \footnote{https://www.zephyrproject.org/}. In the following, we detail our approach and the functions we implemented in the BLE Mesh stack to enable QoS in BLE Mesh networks.

\subsection{Adding the priority class to mesh packets}
\label{addingpriority}

The proposed method enables QoS by adding the priority value to the mesh packets every time a model  sends a message. Here, the unique opcode of each model is used to define the packet's priority. This opcode is sent as an argument through the access and transport layer functions to the network layer, where the opcode is translated to a priority class of 1-byte size, allowing a maximum of 256 priority classes.

\begin{figure}[t]
\centering
\includegraphics[scale=0.175]{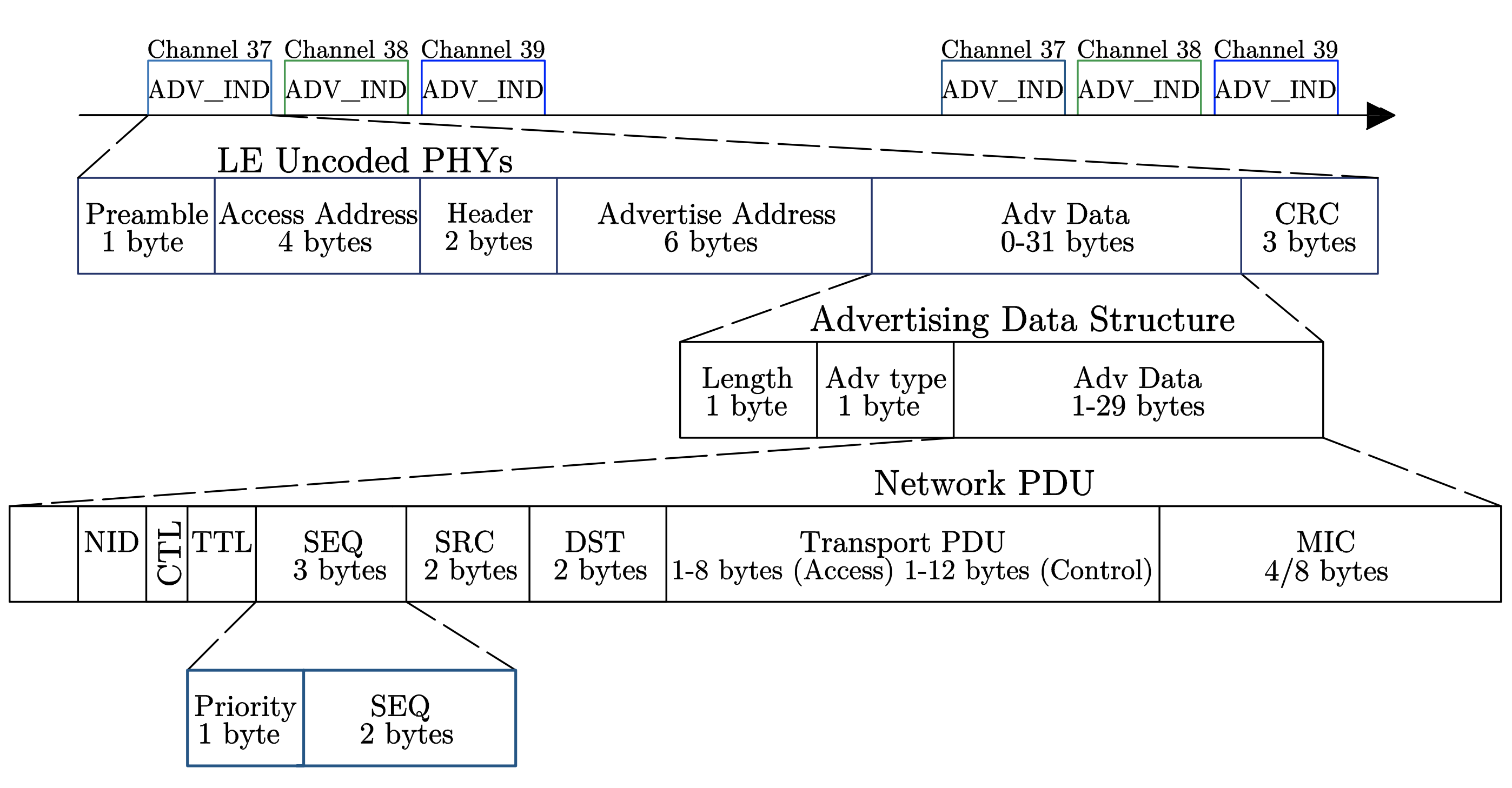}
\vspace{-1em}
\caption{ QoS priority class allocated into the sequence field (SEQ) in the BLE Mesh network-layer protocol data unit (PDU).}

\label{fig:pdu}
\end{figure}

Once the priority value is received on the network layer, it is added as another field of the network PDU. The extra 8 bits of the priority class are allocated into the sequence number (SEQ) field. Here the 24 bits of the SEQ field are reduced to 16, and the remaining 8 bits are assigned to the priority class. Figure \ref{fig:pdu} shows the BLE Mesh network PDU structures with the priority field. It is important to note that the 16 bits assigned to the SEQ number provide enough packet counts to not compromise the security of the BLE Mesh standard against replay attacks \cite{mesh101}. If necessary, the SEQ field size could be increased to 20 bits.

\subsection{Configuring parameters for QoS}

The configurable transmission parameters represent a group of parameters located in the different structures of the BLE Mesh protocol. The parameters can be set to optimize the network performance and give personalized transmission characteristics to BLE Mesh packets. Table \ref{tab1} summarizes each parameter and its range of values.

\begin{table}[t]
\vspace{-0.2em}
\caption{BLE Mesh transmission parameters}
\centering
\fontsize{8}{10}\selectfont 
\begin{tabular}{|p{2.1cm}|p{3.2cm}|p{1.8cm}|}
\hline
\textbf{Parameter}                         & \textbf{Description}                                               & \textbf{Values}                    \\ \hline
Number of Rebroadcast Repetitions (N.Rep) & Number of times a packet is rebroadcast by a network node                                                   & 0 to 1000                 \\ \hline
Advertising Interval  (Adv. Interval)            & Time between advertising intervals                                                                          & 20 ms to 10.24 s          \\ \hline
Time to Live (TTL)                    & The max number of hops a packet can perform                                                               & 0 to 127                  \\ \hline
Transmission Power (Tx Power)               & The transmission power of the BLE radio connection$^{\mathrm{*}}$                                                                & (4, 0, -8, -20, -40 ) dBm \\ \hline
\multicolumn{3}{p{8cm}}{$^{\mathrm{*}}$Changing the transmission power is a feature in the Nordic nRF52832 System-On-Chip.}
\end{tabular}
\label{tab1}
\vspace{-2em}
\end{table}

When a new node is added to the BLE Mesh network, the network transmission parameters, the addresses, and different publish/subscription rights are set during a process called \textit{provisioning}. After this process, only the \textit{number of rebroadcasts}, \textit{advertising interval}, and \textit{TTL} can be changed by executing manual runtime configurations from an external order of the provisioner node. In contrast, the \textit{Transmission Power} cannot be changed during or after the provisioning process. Thus, changing the network parameters from an external order each time a new packet is sent would be effort- and time-inefficient for the source nodes and impossible for the relay nodes. Accordingly, the first step of our solution is to vary these parameters for each packet directly at the nodes, enabling the possibility of having network transmission settings for each packet according to its given priority.

\subsection{Transmitting packets with priority over BLE Mesh networks}

The Packet priority class determines the network transmission parameters for the source and relay nodes in the mesh network. When a packet is transmitted for the first time, the source node elements transmitting it uses a model for every priority. This model sets the initial network transmission parameters and the message opcode according to the priority. As explained in Subsection \ref{addingpriority}, the packet is broadcast with the priority field added at the network layer. For each packet received at a relay node, the network layer determines if this packet is addressed to the node unicast address or to the group address and needs to be relayed to the next node. In the second case, if the packet needs to be relayed, the implemented functions on the network layer determine its priority, set the correct node transmission parameters, and forward the packet to the next node. The BLE Mesh packets without a priority (default) are set with second-priority transmission parameters. Algorithm \ref{alg:qos} summarizes the relay process with our QoS implementation.

\begin{algorithm} [t]
        \small
	\caption{Network-Layer packet relay process with QoS.}
        
	\label{alg:qos}
	\begin{algorithmic}[1]
        \WHILE{True}
        \STATE newPacket $\leftarrow$ receivePacket(\,)
        \STATE packetAddress = getAddress(newPacket)
		\IF{packetAddress == nodeAddress}
        \STATE toTransportLayer(\,)
        \ELSIF{packetAddress == groupAddress}
        \STATE toTransportLayer(\,)
		\IF{toRelay == True}
        \STATE checkPriority(\,)
        \STATE setNetworkPara (\,)
		\STATE relay(\,)
        \ELSE
        \STATE discard(\,)
		\ENDIF
        \ENDIF
        \ENDWHILE
	\end{algorithmic}

\end{algorithm}

\section{Experimental setup and data collection}\label{Sec_IV}

To test the performance of BLE Mesh with the proposed QoS implementation, we deployed an experimental BLE Mesh network. This section presents its implementation and the phases of the experiment campaign. 

\subsection{BLE Mesh experimental network}

\begin{figure}[t]
\centering
\includegraphics[scale=0.19]{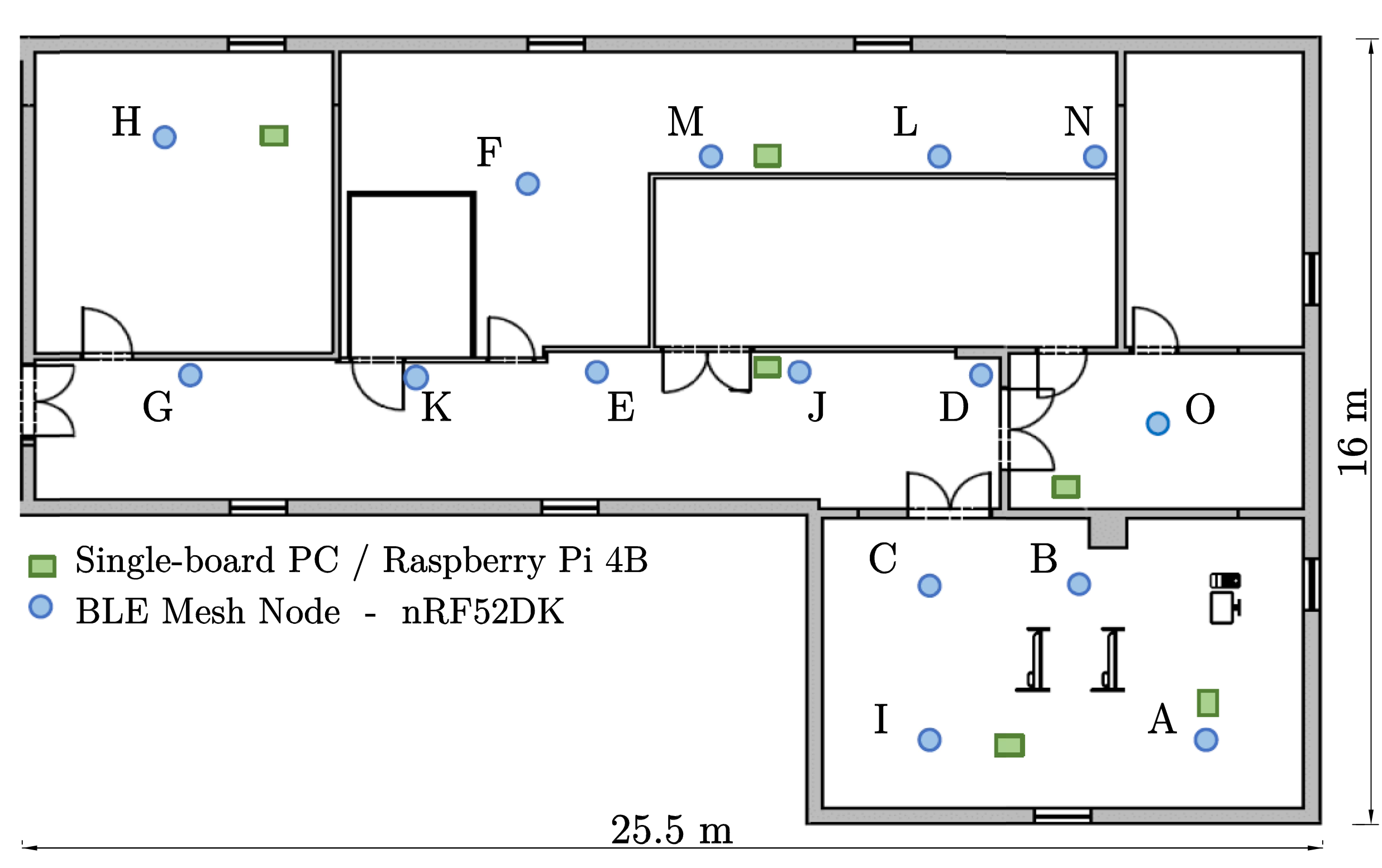}
\caption{BLE Mesh experimental network deployment.}
\label{fig:setup}
\end{figure}

The experimental BLE Mesh network is deployed on a full floor of an office building with a total area of around 400 meters. Here, 15 BLE Mesh nodes are placed in equal distance distribution to represent a smart healthcare sensor network. Every node represents a sensor device generating unsegmented 11-byte BLE Mesh data packets and pushing them in the network. This network setup comprises 15 Nordic Semiconductor nRF52DK development kits with the nRF52832 System-On-Chip\footnote{https://www.nordicsemi.com/products/nrf52832}. The nodes are programmed using version 1.8.0 of the nRF Connect SDK with Zephyr RTOS real-time operating system and version 5.2 of the Bluetooth core specification. For the data collection process, 6 Raspberry Pi 4B\footnote{https://www.raspberrypi.com/products/raspberry-pi-4-model-b/} (RPi) single-board computers are placed next to the nodes. The RPis are synchronized using Network-Time-Protocol. They collect the nodes' information through a serial connection and upload it to a server. Figure \ref{fig:setup} shows the experimental network.

The nodes are provisioned and set with the relay feature, enabling them to send, receive and relay messages. Each node has three elements, each one with a unicast address. Each element represents a sensor in our experiments, and each of the three sensors will have a different network configuration. This way, the source nodes provide three priority classes, one for each sensor. Table \ref{tab:priority} summarizes the network transmission characteristics for each priority.

\begin{table}[t]
\vspace{1em}
\caption{BLE Mesh QoS priority list}
\centering
\small
\begin{tabular}{|p{1.7cm}|p{0.9cm}|p{0.6cm}|p{1.7cm}|p{1.2cm}|} 
\hline
\textbf{Priority} &\textbf{N. Rep} & \textbf{TTL} & \textbf{Adv. Interval~} & \textbf{Tx Power}  \\ \hline\hline
1 - Element 1 - Sensor 1 & 2$^{\mathrm{*}}$                                  & 7                           & 20 ms                          & 4 dBm                        \\ 
\hline
2 - Element 2 - Sensor 2 & 2$^{\mathrm{*}}$                                   & 5                           & 100 ms                         & -8 dBm                       \\ 
\hline
3 - Element 3 - Sensor 3 & 2$^{\mathrm{*}}$                                 & 3                           & 200 ms                         & -20 dBm                      \\
\hline
\multicolumn{5}{p{8cm}}{$^{\mathrm{*}}$The BLE Mesh standard recommends setting the \textit{Number of Rebroadcasts (N.rep)} to 2 \cite{Woolley2020reli}.}
\end{tabular}
\label{tab:priority}
\vspace{-1.2em}
\end{table}

\subsection{Experiment campaign}

Our experiment campaign consists of two experiments. For every experiment, the data is collected by a Raspberry Pi at the source and the destination nodes. Table \ref{Tab_data} illustrates the structure of the database table. The two experiments are detailed as follows:
\begin{enumerate}
    \item \textit{Experiment 1: Single-channel scenario:}\\
    This experiment aims to evaluate the BLE Mesh network performance in a limited congestion network scenario. Therefore, the designated source node is A, the destination node is H, and the other 13 mesh nodes work as receivers and relays (See Figure \ref{fig:setup}). A sends to H 6000 unsegmented data packets with random priorities with a packet generation rate of 2 seconds.
    \item \textit{Experiment 2: Multi-channel scenario:}\\
    Experiment two aims to evaluate the BLE Mesh network performance in a relatively high congestion network scenario. To this end, we defined 2 traffic channels. The first channel's source node is A, and the destination node is H. The second channel's source node is N, and the destination node is G (See Figure \ref{fig:setup}). The other 11 nodes work as receivers and relays. Each source node sends 6000 unsegmented data packets with random priorities and a generation rate of 2 seconds.
\end{enumerate}

\begin{table*}[t]
\caption{BLE Mesh experiment dataset structure}
\centering
\label{Tab_data}
\small
\begin{tabular}{|l|l|l|l|l|l|l|l|l|l|l|} 
\hline
\textbf{Timestamp (ms)} & \textbf{Test Id} & \textbf{Packet Id} & \begin{tabular}[c]{@{}l@{}}\textbf{Sender}\\\textbf{Address}\end{tabular} & \begin{tabular}[c]{@{}l@{}}\textbf{Receiver }\\\textbf{Address}\end{tabular} & \textbf{TTL} & \begin{tabular}[c]{@{}l@{}}\textbf{Tx Power~}\\\textbf{(dBm)}\end{tabular} & \begin{tabular}[c]{@{}l@{}}\textbf{Priority}\\\textbf{Class}\end{tabular} & \textbf{Delivered$^{\mathrm{*}}$} & \begin{tabular}[c]{@{}l@{}}\textbf{Number}\\\textbf{of hops$^{\mathrm{*}}$}\end{tabular} & \textbf{PDT$^{\mathrm{*}}$ (ms)}  \\ 

\hline
1670585825695           & 1                & 18                & 0x0091                                                                    & 0x00C4                                                                       & 7            & 4                                                                          & 1                                                                         & 1                  & 1                                                                         & 17                \\ 
\hline
1670585837784           & 1                & 24                & 0x0091                                                                    & 0x00C4                                                                       & 7            & 4                                                                          & 1                                                                         & 1                  & 0                                                                         & 12                \\ 
\hline
1670585847856           & 1                & 29                & 0x0092                                                                    & 0x00C4                                                                       & 5            & -8                                                                          & 2                                                                         & 1                  & 1                                                                         & 20                \\
\hline
1670585851885           & 1                & 31                & 0x0092                                                                    & 0x00C4                                                                       & 5            & -8                                                                          & 2                                                                         & 1                  & 2                                                                         & 29                \\  
\hline
1670585880137           & 1                & 45                & 0x0093                                                                    & 0x00C4                                                                       & 3            & -20                                                                          & 3                                                                         & 1                  & 2                                                                         & 233                 \\
\hline
1670585884168           & 1                & 47                & 0x0093                                                                    & 0x00C4                                                                       & 3            & -20                                                                          & 3                                                                         & 1                  & 2                                                                         & 227                \\ 
\hline
\multicolumn{11}{p{17cm}}{$^{\mathrm{*}}$The variables \textit{Delivered, Number of hops,} and \textit{PDT (ms)} are variables generated after data collection to evaluate the network performance.}
\end{tabular}
\end{table*}

\section{Results}\label{Sec_V}
This section summarizes the most important insights obtained from our experiment campaign. Three Key Performance Indicators (KPIs) are defined to evaluate the performance of our QoS implementation.
\begin{itemize}
    \item \textit{The Packet Delivery Time (PDT)} measures the elapsed time in milliseconds of a message from the source to the destination.
    \item \textit{The Packet Delivery Rate (PDR)} indicates the delivered packets over the total number of packets sent.
    \item \textit{The Number of Hops (N.Hops)} counts the number of hops that a packet performed to reach its destination.
\end{itemize}
    
\subsection{\textit{Experiment 1: Single-channel scenario.}}

\begin{figure}[t]
    \centering
    \includegraphics[scale=0.41]{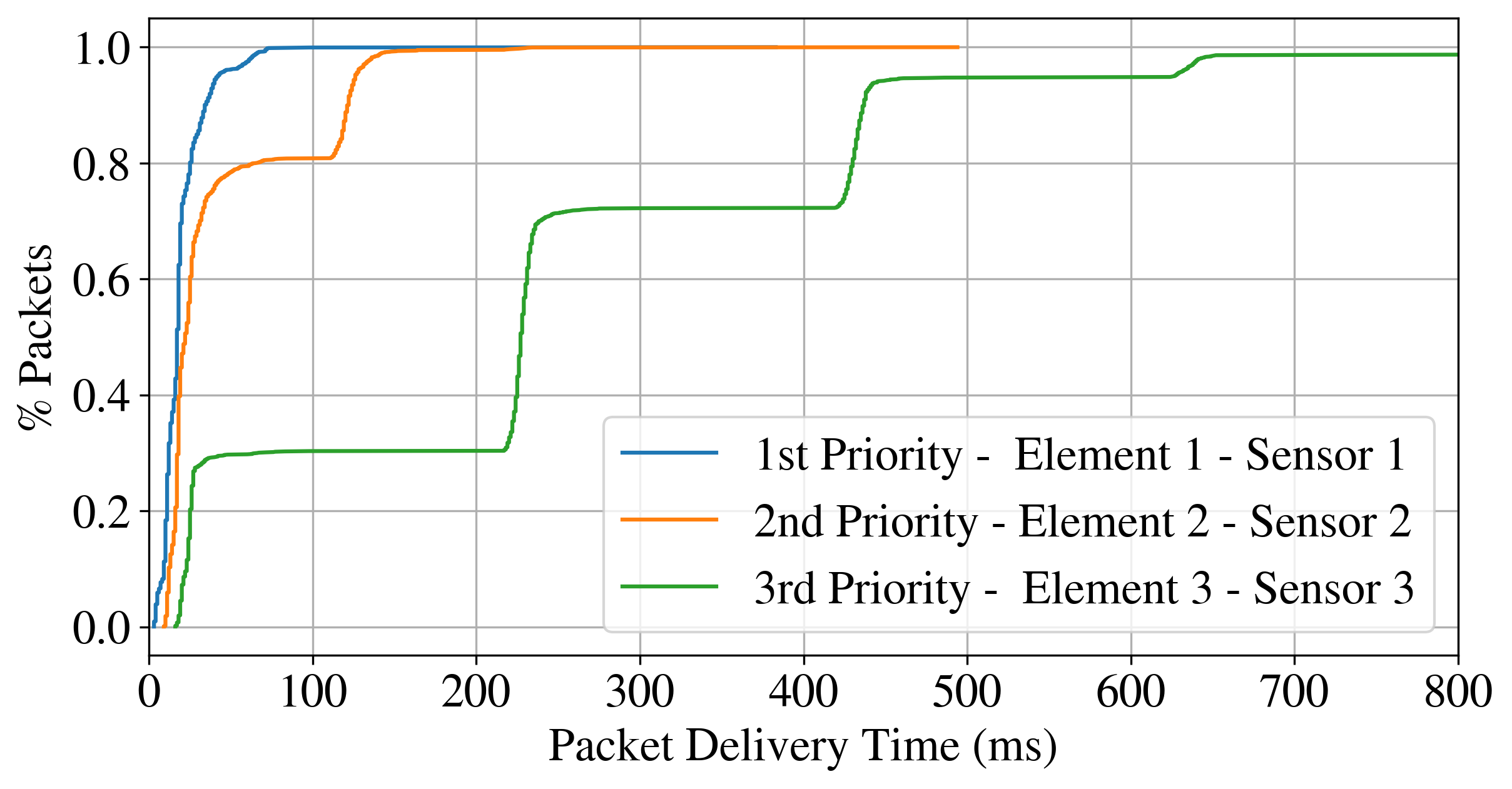}
    \caption{Experiment 1 results for single-channel scenario (A to H).}
    \vspace{-1.5em}
    \label{fig:exp1}

    \end{figure}

In this experiment, the three defined priorities can be clearly distinguished. Figure \ref{fig:exp1} shows the empirical cumulative distribution function (eCDF) for the PDT corresponding to each priority. As shown in the figure, 80\% of the first priority data packets were delivered in less than 20 ms. For the second and third priorities, the same 80\% of the packets were delivered in less than 100 ms and 300 ms, respectively. These results confirm that our QoS implementation shows a clear difference in terms of PDT between the three priorities considered. Table \ref{Tab_exp1} presents the detailed statistics of the experiment results, clearly showing the difference in performance between the three adopted priority classes. It is worth noting that this experiment analyzes the ideal case, in which the traffic is only generated in one channel without interference from other sources or channels. This ideal traffic scenario is also validated with the PDR analysis. As shown in Table \ref{Tab_exp1}, the first and second priorities have a PDR of 1, and the third priority have a PDR of 0.905. The number of hops is also related to the assigned priority, with a minimum average of 1.882 and 3 hops for the first and third priorities, respectively. For the third priority, the PDR (0.905) according to the number of hops can be interpreted as a border effect of the experiment and used to define the correct TTL limit values for each application.
    
    \begin{table}[t]
    \caption{Experiment 1: Key Performance Indicators}
    \label{Tab_exp1}
    \centering
    \small
    \begin{tabular}{|l|l|l|l|} 
    \hline
    \begin{tabular}[c]{@{}l@{}}\textbf{Key Performance }\\\textbf{Indicator}\end{tabular} & \textbf{Priority 1} & \textbf{Priority 2} & \textbf{Priority 3}  \\ 
    \hline \hline
    PDR                                                                                   & 1                   & 1                   & 0.905               \\ 
    \hline\hline
    Number of hops Avg                                                                                   & 1.882                   & 2.495                   & 3               \\ 
    \hline\hline
    PDT Avg (ms)                                                                              & 19.606              & 42.163              & 237.814              \\ 
    \hline
    PDT Std. Dev (ms)                                                                     & 14.742              & 43.693              & 185.764              \\ 
    \hline
    PDT Min (ms)                                                                              & 3                   & 9                   & 16                   \\ 
    \hline
    PDT Max (ms)                                                                              & 383                 & 494                 & 1251                 \\
    \hline
    \end{tabular}

    \end{table}

\subsection{\textit{Experiment 2: Multi-channel scenario.}}

\begin{figure}[t]
    \centering

    \includegraphics[scale=0.41]{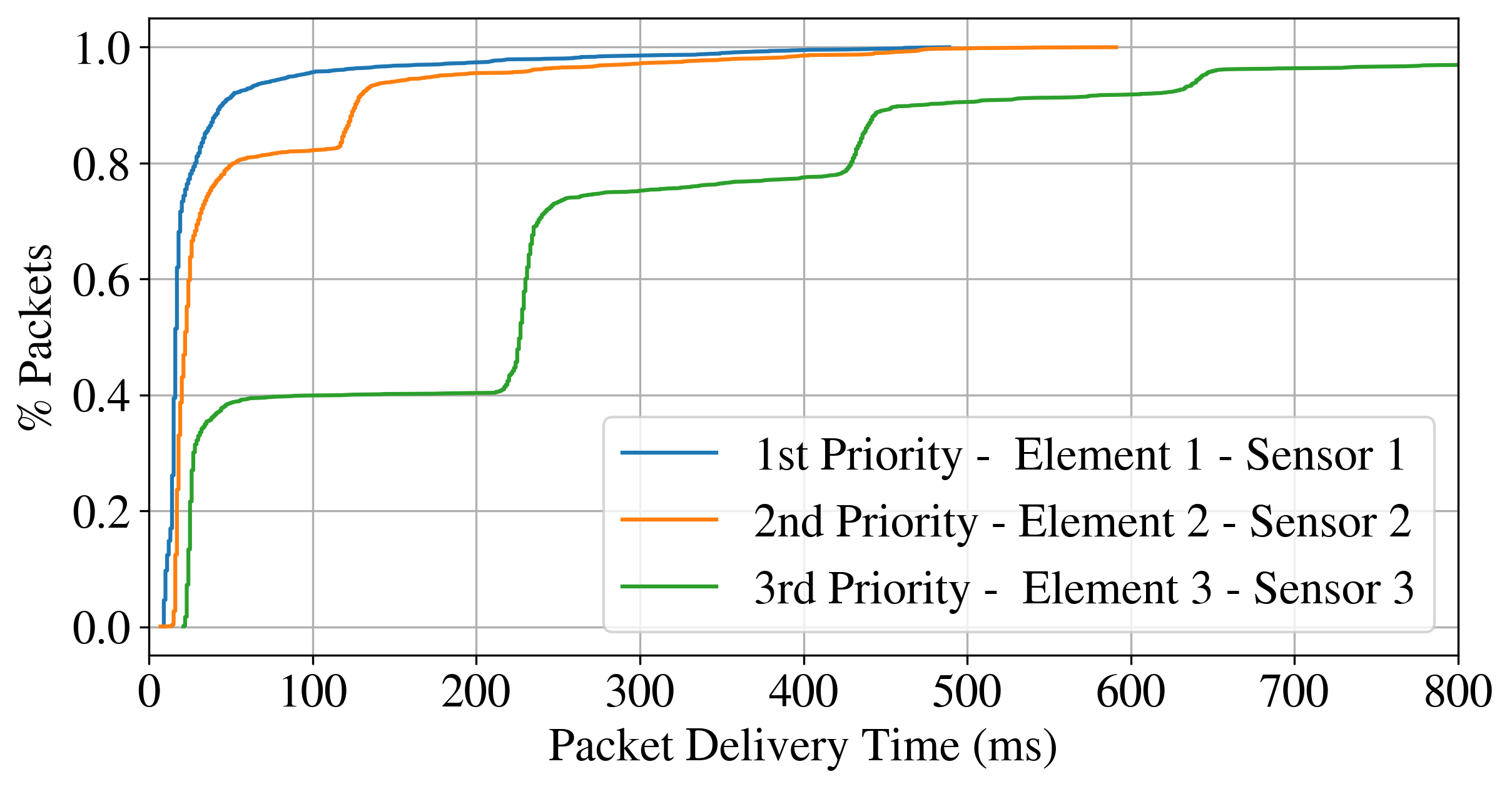}
    \caption{Experiment 2 results for channel 1 (A to H) in a multi-channel scenario.}
    \label{fig:exp21}

    \end{figure}

\begin{figure}[t]
    \centering
    \includegraphics[scale=0.41]{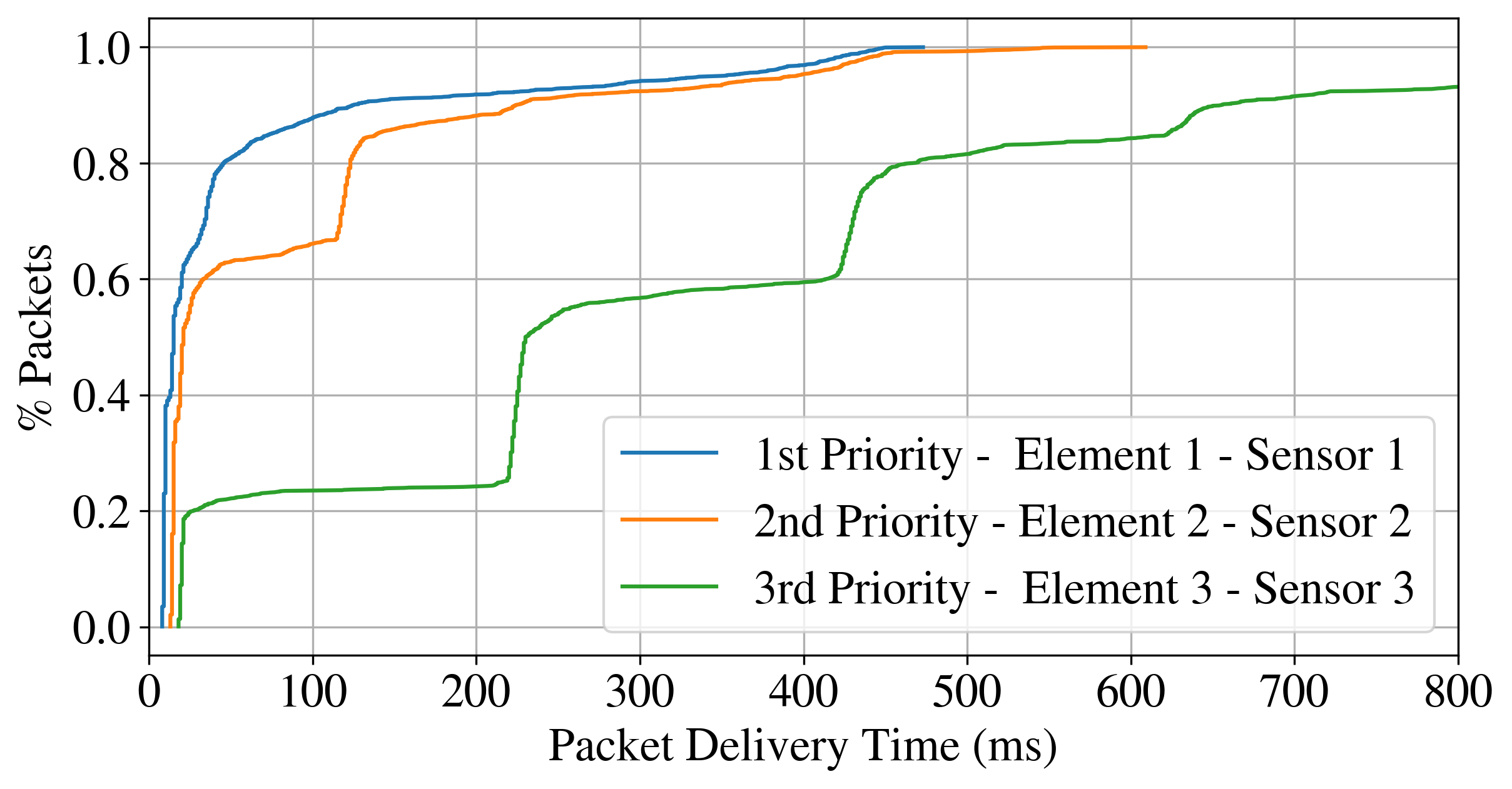}
    \caption{Experiment 2 results for channel 2 (N to G) in a multi-channel scenario.}
    \label{fig:exp22}
    \vspace{-1em}
    \end{figure}

The three defined priorities can be clearly distinguished in this experiment for channel 1 and channel 2. The first important point to note is that the average PDT values are higher for the two channels of this experiment (See Table \ref{tab:exp2}). Second, there are differences in the performance between channels 1 and 2. Figures \ref{fig:exp21} and \ref{fig:exp22} show the PDT eCDFs of both channels, where for the first priority, the difference between channels in average PDT is 22.536 ms, and for the second and third priorities, the difference is 31,133 and 98,821 ms, respectively. Table \ref{tab:exp2} shows the detailed statistics of the second experiment's results. As shown in the table, the PDR remains 1 for both channels' first and second priority. However, for the third priority, channel 2 performs 2,3\% better. For both channels, the number of hops gradually increases depending on the priority, e.g., in channel 2, from 1.643 to 2.939 hops executed by the first and third priority, respectively. The results of experiment 2 also validate the effectiveness of our QoS implementation. It shows the differences between the channels and the differences between priorities in a more realistic network scenario, where more than one channel is exchanging data packets.

    \begin{table}[t]

    \caption{Experiment 2: Key Performance Indicators} 
    \centering
    \small
    \begin{tabular}{|l|l|l|l|} 
    \hline
    \begin{tabular}[c]{@{}l@{}}\textbf{Key Performance}\\\textbf{Indicator}\end{tabular} & \textbf{Priority 1} & \textbf{Priority 2} & \textbf{Priority 3}  \\ 
    \hline\hline
    \multicolumn{4}{|c|}{\textbf{Channel 1 (A to H)}}                                                                                                       \\ 
    \hline\hline
    PDR                                                                                  & 1                   & 1                   & 0.947               \\ 
    \hline\hline
    Number of hops Avg                                                                                   & 1.885                   & 2.187                   & 2.888               \\ 
    \hline\hline
    PDT Avg (ms)                                                                             & 30.579    & 52.392    & 227.004     \\ 
    \hline
    PDT Std. Dev (ms)                                                                    & 54.170    & 77.935    & 221.675     \\ 
    \hline
    PDT Min (ms)                                                                             & 9                   & 7                   & 21                   \\ 
    \hline
    PDT Max (ms)                                                                             & 489                 & 591                 & 1464                 \\ 
    \hline\hline
    \multicolumn{4}{|c|}{\textbf{Channel 2 (N to G)}}                                                                                                       \\ 
    \hline\hline
    PDR                                                                                  & 1                   & 1                   & 0.923               \\ 
    \hline\hline
    Number of hops Avg                                                                                   & 1.643                   & 2.478                   & 2.939               \\ 
    \hline\hline
    PDT Avg (ms)                                                                             & 53.115    & 83.525    & 325.825     \\ 
    \hline
    PDT Std. Dev (ms)                                                                    & 96.850    & 112.398    & 261.839     \\ 
    \hline
    PDT Min (ms)                                                                             & 8                   & 13                  & 18                   \\ 
    \hline
    PDT Max (ms)                                                                             & 473                 & 609                 & 1635                 \\
    \hline
    \end{tabular}
    \label{tab:exp2}

    \end{table}

\section{Conclusions}\label{Sec_VI}

We presented a QoS method for BLE Mesh networks implemented into the network layer of BLE Mesh without interfering with the standard itself. We conducted two extensive measurements to evaluate the performance of the proposed QoS implementation. Both experiments clearly confirmed the differences between the adopted priority classes. We show that a PDR of 1 is always guaranteed for the first and the second traffic classes. The PDT also confirmed the priority differentiation, presenting values of approximately 20 ms and 325 ms for the first and third priority classes, respectively, showing a gap of approximately 300 ms between them. Moreover, the number of hops is also associated with the priorities, varying from 1.8 to 3 hops from the first to the third priority.
Exploring machine-learning-based parameters selection to improve the parameters selection process for the different QoS classes in BLE Mesh networks is a potential future work.

\section{Acknowledgments}
The present work has received funding from the European Union's Horizon 2020 Marie Skłodowska Curie Innovative Training Network Greenedge (GA. No. 953775). The  work  of  Hazem  Sallouha  was  funded  by  the  Research Foundation Flanders (FWO), Postdoctoral Fellowship No. 12ZE222N.

%
%
\balance
\bibliographystyle{ieeetr}
\bibliography{ewsn-full}  
\end{document}